\documentclass{amsart}

\newtheorem{theorem}{Theorem}[section]
\newtheorem{lemma}[theorem]{Lemma}

\theoremstyle{definition}

\newtheorem{example}[theorem]{Example}

\theoremstyle{remark}

\numberwithin{equation}{section}

\begin{document}
	
	\title{Structure and Rank of Cyclic codes over a class of non-chain rings}
	
	
	\author{Nikita Jain} 
	\address{Punjab Engineering College (Deemed to be University), Chandigarh}
	\email{nikitajain.phd19appsc@pec.edu.in}
	\thanks{ Nikita Jain would like to thank Council of Scientific and Industrial Research (CSIR) India, for providing fellowship in support of this research.
	}
	
	\author{Sucheta Dutt}
	\address{Punjab Engineering College (Deemed to be University), Chandigarh}
	\email{sucheta@pec.edu.in}
	
	\author{Ranjeet Sehmi}
	\address{Punjab Engineering College (Deemed to be University), Chandigarh}
	\email{rsehmi@pec.edu.in}
	
	\subjclass[2020]{Primary: 94B15, 20M05, 15A03, 54A25, 13C12 }
	
	\keywords{Cyclic code, Generator, Rank, Cardinality, Rings}
	
	\begin{abstract}
		The rings $Z_{4}+\nu Z_{4}$ have been classified into chain rings and non-chain rings on the basis of the values of $\nu^{2} \in Z_{4}+\nu Z_{4}.$ In this paper, the structure of cyclic codes of arbitrary length over the rings $Z_{4}+\nu Z_{4}$ for those values of $\nu^{2}$ for which these are non-chain rings has been established. A unique form of generators of these codes has also been obtained. Further, rank and cardinality of these codes have been established by finding minimal spanning sets for these codes.
	\end{abstract}
	\maketitle
	\section{Introduction}
	From a mathematical point of view, one of the main aims of algebraic coding theory is to construct codes which are able to detect and correct maximum number of errors during data trasmission. In order to construct such codes, it is important to know the structure of a code.
	
	The class of cyclic codes is one of the significant classes of codes, as these codes offer efficient encoding and decoding of the data using shift registers. These codes have good error detecting and error correcting capabilities. The theory of cyclic codes over finite fields is well established. The study of cyclic codes over rings started after the remarkable work done by Calderbank et al. \cite{10} wherein a Gray map was introduced to show that some non-linear binary codes can be viewed as binary images of linear codes over $Z_{4}$. 
	Recent research involves various approaches to determine the generators of cyclic codes of arbitrary length over various finite commutative rings.  A vast literature is available on cyclic codes over integer residue rings \cite{1,9,11}, Galois rings \cite{12,14} and finite chain rings \cite{7,15}.
	
	The generators of cyclic codes of arbitrary length over finite chain rings of the type $Z_{2}+uZ_{2}, u^{2}=0$ and $Z_{2}+uZ_{2}+u^{2}Z_{2}, u^{3}=0$ have been obtained by  Abualrub and Siap \cite{2}. The same approach is used to find the generators of cyclic codes over the ring $Z_{2}[u]/\left\langle u^{k}\right\rangle$ by Ashker and Hamoudeh \cite{5} and $Z_{p}[u]/\left\langle u^{k}\right\rangle$ by Abhay Kumar and Kewat \cite{17}.The structure of linear and cyclic codes of odd length over a finite non-chain ring $F_{2}[u,v]/\left\langle u^{2},v^{2},uv-vu\right\rangle$ has been determined by Yildiz and Karadeniz \cite{20,21}. A unique set of generators of cyclic codes over the ring $F_{2^{m}}[u,v]/\left\langle u^{2},v^{2},uv-vu\right\rangle$ have been obtained by Sobhani and Molakarimi \cite{18}. The structure of cyclic codes  over the ring $F_{2}[u_{1},u_{2},\cdots,u_{k}]/\left\langle u_{i}^{2},u_{j}^{2},u_{i}u_{j}-u_{j}u_{i}\right\rangle$ have been obtained by Dougherty et al. \cite{8}. The structure of cyclic codes of arbitrary length over the ring $Z_{p}[u,v]/\left\langle u^{2},v^{2},uv-vu\right\rangle$ has been determined by Parmod Kumar Kewat et al. \cite{13}.

	Linear and cyclic codes over the non-chain ring $Z_{4}+\nu Z_{4}$, $\nu^{2}=0$ were first introduced by Yildiz et al. \cite{22,19}. The structure of cyclic codes of arbitrary length over  $Z_{4}+\nu Z_{4}$, $\nu^{2}=0$ has been studied by Bandi and Bhaintwal \cite{6}. Cyclic and some constacyclic codes over the non-chain ring $Z_{4}+\nu Z_{4}$, $\nu^{2}=1$ have been studied by Ozen et al. \cite{16}.

	The rings $Z_{4}+\nu Z_{4}$, $\nu^{2} \in Z_{4}+\nu Z_{4}$ have been classified into chain rings and non-chain rings by Adel Alahmadi et al. \cite{4}. They have proved that $Z_{4}+\nu Z_{4}$ is a chain ring for $\nu^{2} \in \{2,3,1+\nu,1+2\nu,1+3\nu,2+2\nu,3+\nu,3+3\nu\}$ and is a non chain ring for $\nu^{2} \in \{0,1,\nu,2\nu,3\nu,2+\nu,2+3\nu,3+2\nu\}.$

	In this paper, a unique form of generators of cyclic codes of arbitrary length over all non-chain rings of the type $Z_{4}+\nu Z_{4}$, $\nu^{2} \in \{0,1,\nu,2\nu,3\nu,2+\nu,2+3\nu,3+2\nu\}$ has been determined. Further, the rank and cardinality of cyclic codes over these rings have been obtained.

	\section{Preliminaries}
	Let $\mathtt{R}$ be a ring with unity. A subset of $\mathtt{R}^{n}$ over $\mathtt{R}$ is called a code of length $n$. A linear code $\mathtt{C}$ of length $n$ is a submodule of $\mathtt{R}^{n}$ over the ring $\mathtt{R}$. An element of a linear code $\mathtt{C}$ is termed as codeword. If for a codeword $(\mathtt{s}_{0},\mathtt{s}_{1},\cdots,\mathtt{s}_{n-1})$ of $\mathtt{C}$, $(\mathtt{s}_{n-1},\mathtt{s}_{0},\cdots,\mathtt{s}_{n-2})$ is also a codeword of $\mathtt{C}$; then $\mathtt{C}$ is called a cyclic code of length $n$ over $\mathtt{R}$. There is a one to one correspondence between the cyclic codes of length $n$ over $\mathtt{R}$ and the ideals of the ring $\mathtt{R}[z]/\left\langle z^{n}-1\right\rangle$. The rank of a cyclic code, denoted by $rank(\mathtt{C}),$ is the number of elements in the minimal (linear) spanning set of the code $\mathtt{C}$ over $\mathtt{R}.$ A finite commutative ring $\mathtt{R}$ is a chain ring if all its ideals form a chain under the inclusion relation; otherwise $\mathtt{R}$ is a non- chain ring.
	
	Throughout this article, we will denote  the set $\{0,1,\nu,2\nu,3\nu,2+\nu,2+3\nu,3+2\nu\}$ by $\mathtt{S}$ and the non-chain ring $Z_{4}+\nu Z_{4}, \nu^{2}=\theta$ by $\mathtt{R}_{_{\theta}}$ for $\theta \in \mathtt{S}.$ Define\\
	\[ k_{_{\theta}}= \begin{cases}
		\nu & ;\theta \in \{0,\nu,2\nu,3\nu\}\\
		1+\nu & ;\theta \in \{1,3+2\nu\}\\
		2+\nu & ;\theta \in \{2+\nu,2+3\nu\}
	\end{cases} \]\\
	
	The following lemma by Abualrub and Siap \cite{3} determines the structure of cyclic codes of arbitrary length over $Z_{4}$.
	\begin{lemma} \cite{3} Let $\mathtt{C}$ be a cyclic code of arbitrary length $n$ over $Z_{4}$. Then $\mathtt{C}=\left\langle g(z)+2p(z),2a(z)\right\rangle,$ where $ g(z),a(z)$ and $p(z)$ are binary polynomials such that $a(z)|g(z)|z^{n}-1$ and either $p(z)=0$ or $a(z)|p(z)\frac{z^{n}-1}{g(z)}$ with deg $a(z)>$ deg $p(z)$.
	\end{lemma}

	\section{Structure of cyclic codes of arbitrary length over $\mathtt{R}_{_{\theta}}, \theta \in \mathtt{S}$}
	The generators of cyclic codes of arbitrary length over $\mathtt{R}_{_{\theta}}$ for $\theta=0$ have been explicitly studied by Bandi and Bhaintwal \cite{6}. In this section, we establish the structure of cyclic codes of arbitrary length $n$ over all non-chain rings $\mathtt{R}_{_{\theta}}$, $\theta \in \mathtt{S}.$
	\begin{theorem}
		Let $\mathtt{C}_{_{\theta}}$ be a cyclic code of arbitrary length $n$ over the ring  $\mathtt{R}_{_{\theta}},\theta \in \mathtt{S}.$ Then $\mathtt{C}_{_{\theta}} =\langle f_{_{\theta_{1}}}(z),f_{_{\theta_
				{2}}}(z),f_{_{\theta_{3}}}(z),f_{_{\theta_{4}}}(z)\rangle$, where $f_{_{\theta_{1}}}(z)= f_{_{11}}(z)+2f_{_{12}}(z)+k_{_{\theta}}f_{_{13}}(z)+2k_{_{\theta}}f_{_{14}}(z)$, $f_{_{\theta_{2}}}(z)=2f_{_{22}}(z)+k_{_{\theta}}f_{_{23}}(z)+2k_{_{\theta}}f_{_{24}}(z)$, $f_{_{\theta_{3}}}(z)=k_{_{\theta}}f_{_{33}}(z)+2k_{_{\theta}}f_{_{34}}(z)$, $f_{_{\theta_{4}}}(z)=2k_{_{\theta}}f_{_{44}}(z)$ such that the polynomials $f_{_{ij}}(z)$ are in $Z_{2}[z]/{\left\langle z^{n}-1\right\rangle}$ for $1 \leq i \leq 4, i \leq j \leq 4.$ Further,
		\begin{equation}
			f_{_{22}}(z)|f_{_{11}}(z)|z^{n}-1,   
		\end{equation}
		\begin{equation}
			\text{ either}~~ f_{_{12}}(z)=0 \text{ or}~f_{_{22}}(z)|f_{_{12}}(z)\frac{z^{n}-1}{f_{_{11}}(z)} \text{ with deg} ~f_{_{22}}(z)> \text{ deg}~f_{_{12}}(z), 
		\end{equation}
		\begin{equation}
			f_{_{44}}(z)|f_{_{33}}(z)|z^{n}-1,
		\end{equation}
		\begin{equation}
			\text{ either}~~~f_{_{34}}(z)=0  \text{ or}	~f_{_{44}}(z)|f_{_{34}}(z)\frac{z^{n}-1}{f_{_{33}}(z)} \text{ with deg} ~f_{_{44}}(z)> \text{ deg}~f_{_{34}}(z). 
		\end{equation}
	\end{theorem}
	\begin{proof}
		Let $\mathtt{C}_{_{\theta}}$ be a cyclic code of length $n$ over $\mathtt{R}_{_{\theta}}$, $\theta \in \mathtt{S}.$ Define $\phi_{_{\theta}}:\mathtt{R}_{_{\theta}} \rightarrow Z_{4}$ by $\phi_{_{\theta}}(x) =x~ (mod~ k_{_{\theta}}).$ It is easy to see that the maps $\phi_{_{\theta}},$ $\theta \in \mathtt{S}$ are ring homomorphisms. Let $ker_{_{\theta}}=\{x \in \mathtt{C}_{_{\theta}}$ such that $\phi_{_{\theta}}(x)=0\}.$ Clearly, $\phi_{_{\theta}}(\mathtt{C}_{_{\theta}})$ is a cyclic code of length $n$ over $Z_{4}.$ Using Lemma 2.1, we get
		
		$\phi_{_{\theta}}(\mathtt{C}_{_{\theta}})$ = $\left\langle f_{_{11}}(z)+2f_{_{12}}(z),2f_{_{22}}(z)\right\rangle$, where $f_{_{22}}(z)|f_{_{11}}(z)|z^{n}-1$ and \\ either $f_{_{12}}(z)=0$ or $f_{_{22}}(z)|f_{_{12}}(z)\frac{z^{n}-1}{f_{_{11}}(z)}$ with deg $f_{_{22}}(z)>$ deg $f_{_{12}}(z).$ 
		
		Also, $ker_{_{\theta}}$ is $k_{_{\theta}}$ times a cyclic code of length $n$ over $Z_{4}.$ Again using Lemma 2.1, we get $ker_{_{\theta}}$ = $k_{_{\theta}} \left\langle f_{_{33}}(z)+2f_{_{34}}(z),2f_{_{44}}(z)\right\rangle$, where  $f_{_{44}}(z)|f_{_{33}}(z)|z^{n}-1$ and either $f_{_{34}}(z)=0$ or $f_{_{44}}(z)|f_{_{34}}(z)\frac{z^{n}-1}{f_{_{33}}(z)}$ with deg $f_{_{44}}(z)>$ deg $f_{_{34}}(z).$

		It follows that
		$\mathtt{C}_{_{\theta}} =\left\langle f_{_{\theta_{1}}}(z),f_{_{\theta_
				{2}}}(z),f_{_{\theta_{3}}}(z),f_{_{\theta_{4}}}(z)\right\rangle$, where $f_{_{\theta_{1}}}(z)= f_{_{11}}(z)+2f_{_{12}}(z)+k_{_{\theta}}f_{_{13}}(z)+2k_{_{\theta}}f_{_{14}}(z)$, $f_{_{\theta_{2}}}(z)=2f_{_{22}}(z)+k_{_{\theta}}f_{_{23}}(z)+2k_{_{\theta}}f_{_{24}}(z)$, $f_{_{\theta_{3}}}(z)=k_{_{\theta}}f_{_{33}}(z)+2k_{_{\theta}}f_{_{34}}(z)$, $f_{_{\theta_{4}}}(z)=2k_{_{\theta}}f_{_{44}}(z)$ such that the polynomials $f_{_{ij}}(z)$ are in $Z_{2}[z]/{\left\langle z^{n}-1\right\rangle}$ for $1 \leq i \leq 4, i \leq j \leq 4$ and satisfy the conditions (3.1)-(3.4). 
	\end{proof}
	
	Let $\mathtt{C}_{_{\theta}}$ be a cyclic code of length $n$ over $\mathtt{R}_{_{\theta}}$, $\theta \in \mathtt{S},$ generated by the polynomials $f_{_{\theta_{1}}}(z),f_{_{\theta_{2}}}(z),f_{_{\theta_{3}}}(z),f_{_{\theta_{4}}}(z)$ as obtained in Theorem 1. Define Residue and Torsion of $\mathtt{C}_{_{\theta}}$ as\\\\
	Res($\mathtt{C}_{_{\theta}})$= $\Biggl\{a(z)\in \frac{Z_{4}[z]}{\left\langle z^{n}-1\right\rangle}:a(z)+k_{_{\theta}} b(z) \in \mathtt{C}_{_{\theta}}$ for some $b(z) \in \frac{Z_{4}[z]}{\left\langle z^{n}-1 \right\rangle}\Biggr\}$\\
	Tor($\mathtt{C}_{_{\theta}})$=$\Biggl\{a(z)\in \frac{Z_{4}[z]}{\left\langle z^{n}-1 \right\rangle}: k_{_{\theta}} a(z)\in \mathtt{C}_{_{\theta}}\Biggr\}$\\
	Clearly, Res($\mathtt{C}_{_{\theta}}$) and Tor($\mathtt{C}_{_{\theta}})$ are the ideals of the ring $\frac{Z_{4}[z]}{\left\langle z^{n}-1 \right\rangle}$.\\
	Also, define 
	\\
	$\mathtt{C}_{_{\theta_{1}}}$=Res(Res($\mathtt{C}_{_{\theta}}$))= $\mathtt{C}_{_{\theta}}$ mod $(2,k_{_{\theta}})$\\
	$\mathtt{C}_{_{\theta_{2}}}$=Tor(Res($\mathtt{C}_{_{\theta}}$))= $\{a(z)\in Z_{2}[z]:2a(z) \in \mathtt{C}_{_{\theta}}$ mod $k_{_{\theta}}\}$\\
	$\mathtt{C}_{_{\theta_{3}}}$=Res(Tor($\mathtt{C}_{_{\theta}}$))= $\{a(z)\in Z_{2}[z]:k_{_{\theta}}a(z) \in \mathtt{C}_{_{\theta}}$ mod $2k_{_{\theta}}\}$\\
	$\mathtt{C}_{_{\theta_{4}}}$=Tor(Tor($\mathtt{C}_{_{\theta}}$))= $\{a(z)\in Z_{2}[z]:2k_{_{\theta}}a(z) \in \mathtt{C}_{_{\theta}}\}$\\

	It is easy to see that $\mathtt{C}_{_{\theta_{1}}}$,$\mathtt{C}_{_{\theta_{2}}}$,$\mathtt{C}_{_{\theta_{3}}}$,$\mathtt{C}_{_{\theta_{4}}}$  are ideals of the ring $Z_{2}[z]/\left\langle z^{n}-1 \right\rangle$ generated by the unique minimal degree polynomials $f_{_{11}}(z),f_{_{22}}(z),f_{_{33}}(z),f_{_{44}}(z)$ respectively as defined in Theorem 3.1.\\

	\begin{theorem}Let $\mathtt{C}_{_{\theta}} = \langle f_{_{\theta_{1}}}(z),f_{_{\theta_{2}}}(z),f_{_{\theta_{3}}}(z),f_{_{\theta_{4}}}(z)\rangle$ be a cyclic code of arbitrary length $n$ over the ring $\mathtt{R}_{_{\theta}}, \theta \in \mathtt{S};$ where  $f_{_{\theta_{i}}}(z),$ $1 \leq i \leq 4$ are polynomials as defined in Theorem 3.1. Then there exists a set of generators  $\{g_{_{\theta_{1}}}(z),g_{_{\theta_{2}}}(z),g_{_{\theta_{3}}}(z),g_{_{\theta_{4}}}(z)\}$ of $\mathtt{C}_{_{_\theta}},$ where $g_{_{\theta_{1}}}(z)= g_{_{11}}(z)+2g_{_{12}}(z)+k_{_{\theta}}g_{_{13}}(z)+2k_{_{\theta}}g_{_{14}}(z)$, $g_{_{\theta_{2}}}(z)=2g_{_{22}}(z)+k_{_{\theta}}g_{_{23}}(z)+2k_{_{\theta}}g_{_{24}}(z)$, $g_{_{\theta_{3}}}(z)=k_{_{\theta}}g_{_{33}}(z)+2k_{_{\theta}}g_{_{34}}(z)$, $g_{_{\theta_{4}}}(z)=2k_{_{\theta}}g_{_{44}}(z)$ such that the polynomials $g_{_{ij}}(z)$ are in $Z_{2}[z]/{\left\langle z^{n}-1 \right\rangle}$ satisfy the conditions (3.1)-(3.4) as defined in Theorem 1 and $g_{_{ii}}(z)$ are unique minimal degree polynomial generators of $\mathtt{C}_{_{\theta_{i}}}, 1 \leq i \leq 4.$ Also, either $g_{_{ij}}(z)=0$ or deg $g_{_{ij}}(z)<$ deg $g_{_{jj}}(z)$ for $ 1\leq i \leq 3, i<j\leq 4.$ \\
	\end{theorem}
	\begin{proof} Clearly, $f_{_{\theta_{1}}}(z)= f_{_{11}}(z)+2f_{_{12}}(z)+k_{_{\theta}}f_{_{13}}(z)+2k_{_{\theta}}f_{_{14}}(z)$, $f_{_{\theta_{2}}}(z)=2f_{_{22}}(z)+k_{_{\theta}}f_{_{23}}(z)+2k_{_{\theta}}f_{_{24}}(z)$, $f_{_{\theta_{3}}}(z)=k_{_{\theta}}f_{_{33}}(z)+2k_{_{\theta}}f_{_{34}}(z)$, $f_{_{\theta_{4}}}(z)=2k_{_{\theta}}f_{_{44}}(z)$ are the generators of $\mathtt{C}_{_{\theta}}$ such that either $f_{_{12}}=0$ or deg $f_{_{12}} <$ deg $f_{_{22}}$ and either $f_{_{34}}=0$ or deg $f_{_{34}} <$ deg $f_{_{44}}.$ Further, if either $f_{_{ij}}=0$ or deg $f_{_{ij}} <$ deg $f_{_{jj}}$ for all $1 \le i \le 2, i < j \le 4,$ then we get the required result. Otherwise, let us suppose that deg $f_{_{ij}} \geq$ deg $f_{_{jj}}$ for some $ i=1,2$ and $j=3,4.$ Assume that deg $f_{_{ij}} \geq$ deg $f_{_{jj}}$ for (say) $ i=1$ and $j=3,4$ i.e.,  deg $f_{_{13}} \geq$ deg $f_{_{33}}.$ Thus by division algorithm, there exist some $q_{_{13}}(z)$ and $g_{_{13}}(z) \in Z_{2}[z]$ such that $f_{_{13}}(z)=q_{_{13}}(z)f_{_{33}}(z)+g_{_{13}}(z),$ where either $g_{_{13}}(z)=0$ or deg $g_{_{13}}(z) <$ deg $f_{_{33}}(z).$ Consider, $f_{_{\theta_{1}}}(z)-q_{_{13}}(z)f_{_{\theta_{3}}}(z)=f_{_{11}}(z)+2f_{_{12}}(z)+k_{_{\theta}}g_{_{13}}(z)+2k_{_{\theta}}(f_{_{14}}(z)-q_{_{13}}(z)f_{_{34}}(z)).$  Further, deg $(f_{_{14}}(z)-q_{_{13}}(z)f_{_{34}}(z)) \geq $ deg $f_{_{44}}(z),$ then again by division algorithm, there exist some $q_{_{14}}(z)$ and $g_{_{14}}(z)$ such that $f_{_{14}}(z)-q_{_{13}}(z)f_{_{34}}(z)=f_{_{44}}(z)q_{_{14}}(z)+g_{_{14}}(z),$ where either $g_{_{14}}(z)=0$ or deg $g_{_{14}}(z) <$ deg $f_{_{44}}(z).$ Now consider, $f_{_{\theta_{1}}}(z)-q_{_{13}}(z)f_{_{\theta_{3}}}(z)-q_{_{14}}(z)f_{_{\theta_{4}}}(z)=f_{_{11}}(z)+2f_{_{12}}(z)+k_{_{\theta}}g_{_{13}}(z)+2k_{_{\theta}}g_{_{14}}(z).$ Therefore, there exist a polynomial $g_{_{\theta_{1}}}(z)=f_{_{11}}(z)+2f_{_{12}}(z)+k_{_{\theta}}g_{_{13}}(z)+2k_{_{\theta}}g_{_{14}}(z) \in \mathtt{C}_{_{\theta}}$ such that either $g_{_{13}}(z)=0$ or deg $g_{_{13}}(z) <$ deg $f_{_{33}}(z)$ and either $g_{_{14}}(z)=0$ or deg $g_{_{14}}(z) <$ deg $f_{_{44}}(z).$ Also, since $g_{_{\theta_{1}}}(z)$ is a linear combination of $f_{_{\theta_{1}}}(z),f_{_{\theta_{3}}}(z),f_{_{\theta_{4}}}(z),$ we have $\mathtt{C}_{_{\theta}} = \left\langle f_{_{\theta_{1}}}(z),f_{_{\theta_{2}}}(z),f_{_{\theta_{3}}}(z),f_{_{\theta_{4}}}(z)\right\rangle$=$ \left\langle g_{_{\theta_{1}}}(z),f_{_{\theta_{2}}}(z),f_{_{\theta_{3}}}(z),f_{_{\theta_{4}}}(z)\right\rangle.$ Further, if deg $f_{_{ij}}(z) \geq$ deg $f_{_{jj}}(z)$ for other values of $i$ and $j$ also, then we obtain the required set of generators by using the same arguments as above.
	\end{proof}

	In the following theorem, a unique form of the generators of a cyclic code $\mathtt{C}_{_{\theta}}$ of arbitrary length $n$ over $\mathtt{R}_{_{\theta}}, \theta \in \mathtt{S},$ has been determined.

	\begin{theorem} Let $\mathtt{C}_{_{\theta}}=\langle g_{_{\theta_{1}}}(z),g_{_{\theta_{2}}}(z),g_{_{\theta_{3}}}(z),g_{_{\theta_{4}}}(z)\rangle$ be a cyclic code of arbitrary length $n$ over the ring $\mathtt{R}_{_{\theta}}, \theta \in \mathtt{S}$, where $g_{_{\theta_{1}}}(z)= g_{_{11}}(z)+2g_{_{12}}(z)+k_{_{\theta}}g_{_{13}}(z)+2k_{_{\theta}}g_{_{14}}(z)$, $g_{_{\theta_{2}}}(z)=2g_{_{22}}(z)+k_{_{\theta}}g_{_{23}}(z)+2k_{_{\theta}}g_{_{24}}(z)$, $g_{_{\theta_{3}}}(z)=k_{_{\theta}}g_{_{33}}(z)+2k_{_{\theta}}g_{_{34}}(z)$, $g_{_{\theta_{4}}}(z)=2k_{_{\theta}}g_{_{44}}(z)$ such that the polynomials $g_{_{ij}}(z)$ are in $Z_{2}[z]/{\left\langle z^{n}-1 \right\rangle}$ and satisfy the conditions (3.1)-(3.4) as defined in Theorem 3.1 with either $g_{_{ij}}(z)=0$ or deg $g_{_{ij}}(z)<$ deg $g_{_{jj}}(z)$ for $ 1\leq i \leq 3, i<j\leq 4$ and $g_{_{ii}}(z)$ are the unique minimal degree polynomial generators of $\mathtt{C}_{_{\theta_{i}}}, 1 \leq i \leq 4.$ Then the polynomials $g_{_{\theta_{1}}}(z),g_{_{\theta_{2}}}(z),g_{_{\theta_{3}}}(z),g_{_{\theta_{4}}}(z)$ are uniquely determined.
	\end{theorem}
	\begin{proof}
		Consider another set of generators  $\{h_{_{\theta_{1}}}(z),h_{_{\theta_{2}}}(z),h_{_{\theta_{3}}}(z),h_{_{\theta_{4}}}(z)\}$ of $\mathtt{C}_{_{\theta}}$, where $h_{_{\theta_{1}}}(z)= h_{_{11}}(z)+2h_{_{12}}(z)+k_{_{\theta}}h_{_{13}}(z)+2k_{_{\theta}}h_{_{14}}(z)$, $h_{_{\theta_{2}}}(z)=2h_{_{22}}(z)+k_{_{\theta}}h_{_{23}}(z)+2k_{_{\theta}}h_{_{24}}(z)$, $h_{_{\theta_{3}}}(z)=k_{_{\theta}}h_{_{33}}(z)+2 k_{_{\theta}}h_{_{34}}(z),$ $h_{_{\theta_{4}}}(z)=2k_{_{\theta}}h_{_{44}}(z)$ such that the polynomials  $h_{_{ij}}(z)$ are in $Z_{2}[z]/{\left\langle z^{n}-1 \right\rangle}$ and satisfy the conditions (3.1)-(3.4) as defined in Theorem 3.1 with either $h_{_{ij}}(z)=0$ or deg $h_{_{ij}}(z)<$ deg $h_{_{jj}}(z)$ for $ 1\leq i \leq 3, i<j \leq 4$ and $h_{_{{ii}}}(z)$ are the unique minimal degree polynomial generators of $\mathtt{C}_{_{\theta_{i}}}, 1 \leq i \leq 4.$ 
		
		Clearly, $g_{_{ii}}(z)=h_{_{ii}}(z),$ for $1 \leq i \leq 4.$
		Consider,	$g_{_{\theta_{1}}}(z)-h_{_{\theta_{1}}}(z)=2(g_{_{12}}(z)-h_{_{12}}(z))+k_{_{\theta}}(g_{_{13}}(z)-h_{_{13}}(z))+2k_{_{\theta}}(g_{_{14}}(z)-h_{_{14}}(z)) \in \mathtt{C}_{_{\theta}}.$
		This implies that $g_{_{12}}(z)-h_{_{12}}(z) \in \mathtt{C}_{_{\theta_{2}}} =\left\langle g_{_{22}}(z)\right\rangle.$ Also deg $(g_{_{12}}(z)-h_{_{12}}(z))<$ deg $g_{_{22}}(z)$, which is a contradiction because $g_{_{22}}(z)$ is a minimal degree poynomial in  $\mathtt{C}_{_{\theta_{2}}}$. Hence, $g_{_{12}}(z)=h_{_{12}}(z).$ It follows that $g_{_{\theta_{1}}}(z)-h_{_{\theta_{1}}}(z)= k_{_{\theta}}(g_{_{13}}(z)-h_{_{13}}(z))+2k_{_{\theta}}(g_{_{14}}(z)-h_{_{14}}(z)) \in \mathtt{C}_{_{\theta}}$
		which implies that $g_{_{13}}(z)-h_{_{13}}(z) \in \mathtt{C}_{_{\theta_{3}}}=\left\langle g_{_{33}}(z)\right\rangle.$
		As deg $(g_{_{13}}(z)-h_{_{13}}(z)) <$ deg $g_{_{33}}(z)$, we must have $g_{_{13}}(z)=h_{_{13}}(z)$.

		Subsequently, $g_{_{\theta_{1}}}(z)-h_{_{\theta_{1}}}(z)=2k_{_{\theta}}(g_{_{14}}(z)-h_{_{14}}(z)) \in \mathtt{C}_{_{\theta}}$ implying that $g_{_{14}}(z)-h_{_{14}}(z) \in \mathtt{C}_{_{\theta_{4}}} =\left\langle g_{_{44}}(z)\right\rangle.$ This together with the fact that deg $(g_{_{14}}(z)-h_{_{14}}(z))< $ deg $g_{_{44}}(z)$, implies that $g_{_{14}}(z)=h_{_{14}}(z)$.\\ In a similar manner, we can prove that $g_{_{23}}(z)=h_{_{23}}(z)$, $g_{_{24}}(z)=h_{_{24}}(z)$ and $g_{_{34}}(z)=h_{_{34}}(z)$. This proves the uniqueness of the polynomials $g_{_{\theta_{1}}}(z),g_{_{\theta_{2}}}(z),g_{_{\theta_{3}}}(z),g_{_{\theta_{4}}}(z).$ 
	\end{proof}
	
	\begin{theorem} Let $\mathtt{C}_{_{\theta}}=\langle g_{_{\theta_{1}}}(z),g_{_{\theta_{2}}}(z),g_{_{\theta_{3}}}(z),g_{_{\theta_{4}}}(z)\rangle,$ be a cyclic code of arbitrary length $n$ over the ring $\mathtt{R}_{_{\theta}}, \theta \in \mathtt{S},$ where the generators $g_{_{\theta_{1}}}(z)= g_{_{11}}(z)+2g_{_{12}}(z)+k_{_{\theta}}g_{_{13}}(z)+2k_{_{\theta}}g_{_{14}}(z)$, $g_{_{\theta_{2}}}(z)=2g_{_{22}}(z)+k_{_{\theta}}g_{_{23}}(z)+2k_{_{\theta}}g_{_{24}}(z)$, $g_{_{\theta_{3}}}(z)=k_{_{\theta}}g_{_{33}}(z)+2k_{_{\theta}}g_{_{34}}(z)$, $g_{_{\theta_{4}}}(z)=2k_{_{\theta}}g_{_{44}}(z)$ are in the unique form as given by Theorem 3.3. Then the following relations hold for $g_{_{ij}}(z),$ $1 \leq i \leq 4, i \leq j \leq 4$ in $Z_{2}[z]/\left\langle z^{n}-1\right\rangle.$\\
		\begin{itemize}
			\item [(i)] $g_{_{33}}(z)|\dfrac{z^{n}-1}{g_{_{11}}(z)}\Big(g_{_{13}}(z)-\dfrac{g_{_{12}}(z)}{g_{_{22}}(z)}g_{_{23}}(z)\Big),$\\
			\item [(ii)] $g_{_{44}}(z)|g_{_{23}}(z),$\\
			\item [(iii)] $g_{_{33}}(z)|\dfrac{g_{_{11}}(z)}{g_{_{22}}(z)}g_{_{23}}(z),$\\
			\item [(iv)] $g_{_{44}}(z)|\dfrac{z^{n}-1}{g_{_{22}}(z)}\Big(g_{_{24}}(z)-\dfrac{g_{_{23}}(z)}{g_{_{33}}(z)}g_{_{34}}(z)\Big),$\\
			\item [(v)] 	$g_{_{44}}(z)|g_{_{13}}(z)-\dfrac{g_{_{11}}(z)}{g_{_{22}}(z)}g_{_{24}}(z)+\dfrac{g_{_{11}}(z)}{g_{_{22}}(z)g_{_{33}}(z)}g_{_{23}}(z)g_{_{34}}(z),$\\
			\item [(vi)] $g_{_{44}}(z)|\dfrac{z^{n}-1}{g_{_{11}}(z)}\Big(g_{_{14}}(z)-\dfrac{g_{_{12}}(z)}{g_{_{22}}(z)}g_{_{24}}(z)+\dfrac{-g_{_{13}}(z)+\dfrac{g_{_{12}}(z)g_{_{23}}(z)}{g_{_{22}}(z)}}{g_{_{33}}(z)}g_{_{34}}(z)\Big),$\\
			\item [(vii)]
			$g_{_{33}}(z)|g_{_{11}}(z) ~~  \text{for } \theta \in \{0,1,2\nu,3+2\nu\},$\\
			$g_{_{44}}(z)|g_{_{11}}(z) ~~ \text{for } \theta \in \{0,1,2\nu,3+2\nu\},$\\
			$g_{_{44}}(z)|g_{_{22}}(z) ~~ \text{for } \theta \in \{0,3+2\nu\},$\\
			$g_{_{44}}(z)|g_{_{22}}(z)+g_{_{23}}(z)$ for $\theta \in \{1,2\nu\},$\\
			\item [(viii)] $g_{_{44}}(z)|g_{_{12}}(z)+g_{_{13}}(z)-\dfrac{g_{_{11}}(z)}{g_{_{33}}(z)}g_{_{34}}(z) ~~ \text{for } \theta \in \{1,2\nu\},$\\
			$g_{_{44}}(z)|g_{_{12}}(z)-\dfrac{g_{_{11}}(z)}{g_{_{33}}(z)}g_{_{34}}(z)  ~~\text{for } \theta \in \{0,3+2\nu\},$ \\
			$g_{_{44}}(z)|g_{_{13}}(z) ~~ \text{for } \theta\in \{\nu,3\nu,2+\nu,2+3\nu\}.$
		\end{itemize}
	\end{theorem}
	
	\begin{proof} 
		\begin{itemize}
			\item [(i)] Since $\mathtt{C}_{_{\theta}}$ is an ideal in the $\frac{\mathtt{R}_{_{\theta}}[z]}{\left\langle z^{n}-1 \right\rangle}$, we have
				$\dfrac{z^{n}-1}{g_{_{11}}(z)}\Big(g_{_{11}}(z)+2g_{_{12}}(z)+k_{_{\theta}} g_{_{13}}(z)+2k_{_{\theta}}g_{_{14}}(z)\Big)-\dfrac{z^{n}-1}{g_{_{11}}(z)} \dfrac{g_{_{12}}(z)}{g_{_{22}}(z)} \Big(2g_{_{22}}(z)+k_{_{\theta}}g_{_{23}}(z)+2k_{_{\theta}}g_{_{24}}(z)\Big)$ belongs to $\mathtt{C}_{_{\theta}}.$
			It follows that		$k_{_{\theta}}\dfrac{z^{n}-1}{g_{_{11}}(z)}\Big(g_{_{13}}(z)-\dfrac{g_{_{12}}(z)}{g_{_{22}}(z)}g_{_{23}}(z)\Big)+2k_{_{\theta}}\dfrac{z^{n}-1}{g_{_{11}}(z)}\Big(g_{_{14}}(z)-\dfrac{g_{_{12}}(z)}{g_{_{22}}(z)}g_{_{24}}(z)\Big) \in \mathtt{C}_{_{\theta}},$ which implies that
			$k_{_{\theta}}\dfrac{z^{n}-1}{g_{_{11}}(z)}\Big(g_{_{13}}(z)-\dfrac{g_{_{12}}(z)}{g_{_{22}}(z)}g_{_{23}}(z)\Big)$ belongs to $\mathtt{C}_{_{\theta}} ~~(\text{mod } 2k_{_{\theta}}).$ Hence
			$	\dfrac{z^{n}-1}{g_{_{11}}(z)}\Big(g_{_{13}}(z)-\dfrac{g_{_{12}}(z)}{g_{_{22}}(z)}g_{_{23}}(z)\Big) \in \mathtt{C}_{_{\theta_{3}}} =\left\langle g_{_{33}}(z) \right\rangle$.
			Therefore, $g_{_{33}}(z)|\dfrac{z^{n}-1}{g_{_{11}}(z)}\Big(g_{_{13}}(z)-\dfrac{g_{_{12}}(z)}{g_{_{22}}(z)}g_{_{23}}(z)\Big).$\\
			
			\item [(ii)] Since $2\Big(2g_{_{22}}(z)+k_{_{\theta}} g_{_{23}}(z)+2k_{_{\theta}}g_{_{24}}(z)\Big) \in \mathtt{C}_{_{\theta}}$, we have $2k_{_{\theta}}g_{_{23}}(z) \in \mathtt{C}_{_{\theta}}$. It follows that $g_{_{23}}(z) \in \mathtt{C}_{_{\theta_{4}}} = \left\langle g_{_{44}}(z) \right\rangle,$ and therefore $g_{_{44}}(z)|g_{_{23}}(z).$\\
			
			\item[(iii)] As $2 \Big(g_{_{11}}(z)+2g_{_{12}}(z)+k_{_{\theta}} g_{_{13}}(z)+2k_{_{\theta}}g_{_{14}}(z)\Big)-\dfrac{g_{_{11}}(z)}{g_{_{22}}(z)} \Big(2g_{_{22}}(z)+k_{_{\theta}}g_{_{23}}(z)+2k_{_{\theta}}g_{_{24}}(z)\Big)$ belongs to $\mathtt{C}_{_{\theta}},$ it follows that	$-k_{_{\theta}}\dfrac{g_{_{11}}(z)}{g_{_{22}}(z)}g_{_{23}}(z)  \in \mathtt{C}_{_{\theta}} ~~(\text{mod}~ 2k_{_{\theta}}),$ which implies that
			$\dfrac{g_{_{11}}(z)}{g_{_{22}}(z)}g_{_{23}}(z)  \in \mathtt{C}_{_{\theta_{3}}}=\left\langle g_{_{33}}(z)\right\rangle.$ Therefore,
			$g_{_{33}}(z)|\dfrac{g_{_{11}}(z)}{g_{_{22}}(z)}g_{_{23}}(z).$\\

			\item[(iv)] Since
			$\dfrac{z^{n}-1}{g_{_{22}}(z)}\Big(2g_{_{22}}(z)+k_{_{\theta}}g_{_{23}}(z)+2k_{_{\theta}}g_{_{24}}(z)\Big)- \dfrac{z^{n}-1}{g_{_{22}}(z)}\dfrac{g_{_{23}}(z)}{g_{_{33}}(z)}\Big(k_{_{\theta}}g_{_{33}}(z)+2k_{_{\theta}}g_{_{34}}(z)\Big)$ belongs to $\mathtt{C}_{_{\theta}},$ it follows that	$2k_{_{\theta}}\dfrac{z^{n}-1}{g_{_{22}}(z)}\Big(g_{_{24}}(z)-\dfrac{g_{_{23}}(z)}{g_{_{33}}(z)}g_{_{34}}(z)\Big)$ belongs to $\mathtt{C}_{_{\theta}},$ which implies that $\dfrac{z^{n}-1}{g_{_{22}}(z)}\Big(g_{_{24}}(z)-\dfrac{g_{_{23}}(z)}{g_{_{33}}(z)}g_{_{34}}(z)\Big) \in \mathtt{C}_{_{\theta_{4}}}$.
			Hence, $g_{_{44}}(z)|\dfrac{z^{n}-1}{g_{_{22}}(z)}\Big(g_{_{24}}(z)-\dfrac{g_{_{23}}(z)}{g_{_{33}}(z)}g_{_{34}}(z)\Big).$\\

			\item [(v)] Since $2\Big( g_{_{11}}(z)+2g_{_{12}}(z)+k_{_{\theta}} g_{_{13}}(z)+2k_{_{\theta}}g_{_{14}}(z)\Big)-\dfrac{g_{_{11}}(z)}{g_{_{22}}(z)} \Big(2g_{_{22}}(z)+k_{_{\theta}} g_{_{23}}(z)+2k_{_{\theta}}g_{_{24}}(z)\Big)+ \dfrac{g_{_{11}}(z)}{g_{_{22}}(z)}\dfrac{g_{_{23}}(z)}{g_{_{33}}(z)}\Big(k_{_{\theta}}(g_{_{33}}(z)+2g_{_{34}}(z))\Big) \in \mathtt{C}_{_{\theta}},$ it follows that
			
				$		2k_{_{\theta}}\Big(g_{_{13}}(z)-\dfrac{g_{_{11}}(z)}{g_{_{22}}(z)}g_{_{24}}(z)+\dfrac{g_{_{11}}(z)}{g_{_{22}}(z)}\dfrac{g_{_{23}}(z)}{g_{_{33}}(z)}g_{_{34}}(z)\Big) \in \mathtt{C}_{_{\theta}},$ which implies that 
				$\Big(g_{_{13}}(z)-\dfrac{g_{_{11}}(z)}{g_{_{22}}(z)}g_{_{24}}(z)+\dfrac{g_{_{11}}(z)}{g_{_{22}}(z)}\dfrac{g_{_{23}}(z)}{g_{_{33}}(z)}g_{_{34}}(z)\Big) \in \mathtt{C}_{_{\theta_{4}}}=\left\langle g_{_{44}}(z)\right\rangle.$
			Therefore, $g_{_{44}}(z)|g_{_{13}}(z)-\dfrac{g_{_{11}}(z)}{g_{_{22}}(z)}g_{_{24}}(z)+\dfrac{g_{_{11}}(z)}{g_{_{22}}(z)g_{_{33}}(z)}g_{_{23}}(z)g_{_{34}}(z).$\\\\
			
			\item [(vi)] Since	$\dfrac{z^{n}-1}{g_{_{11}}(z)}\Big(g_{_{11}}(z)+2g_{_{12}}(z)+k_{_{\theta}} g_{_{13}}(z)+2k_{_{\theta}}g_{_{14}}(z)\Big)-\dfrac{z^{n}-1}{g_{_{11}}(z)}\dfrac{g_{_{12}}(z)}{g_{_{22}}(z)}\Big(2g_{_{22}}(z)+k_{_{\theta}} g_{_{23}}(z)+2k_{_{\theta}}g_{_{24}}(z)\Big)+\dfrac{z^{n}-1}{g_{_{11}}(z)}\Big(\dfrac{-g_{_{13}}(z)+\dfrac{g_{_{12}}(z)}{g_{_{22}}(z)}g_{_{23}}(z)}{g_{_{33}}(z)}\Big)\Big(k_{_{\theta}}(g_{_{33}}(z)+2g_{_{34}}(z))\Big)$ belongs to $\mathtt{C}_{_{\theta}},$ it follows that\\
			$2k_{_{\theta}}\dfrac{z^{n}-1}{g_{_{11}}(z)}\Big(g_{_{14}}(z)-\dfrac{g_{_{12}}(z)}{g_{_{22}}(z)}g_{_{24}}(z)+\dfrac{-g_{_{13}}(z)+\dfrac{g_{_{12}}(z)g_{_{23}}(z)}{g_{_{22}}(z)}}{g_{_{33}}(z)}g_{_{34}}(z)\Big) \in \mathtt{C}_{_{\theta}},$ which implies that $ \dfrac{z^{n}-1}{g_{_{11}}(z)}\Big(g_{_{14}}(z)-\dfrac{g_{_{12}}(z)}{g_{_{22}}(z)}g_{_{24}}(z)+\dfrac{-g_{_{13}}(z)+\dfrac{g_{_{12}}(z)g_{_{23}}(z)}{g_{_{22}}(z)}}{g_{_{33}}(z)}g_{_{34}}(z)\Big)$ belongs to $\mathtt{C}_{_{\theta_{4}}}.$ Therefore,\\
			$ g_{_{44}}(z)|\dfrac{z^{n}-1}{g_{_{11}}(z)}\Big(g_{_{14}}(z)-\dfrac{g_{_{12}}(z)}{g_{_{22}}(z)}g_{_{24}}(z)+\dfrac{-g_{_{13}}(z)+\dfrac{g_{_{12}}(z)g_{_{23}}(z)}{g_{_{22}}(z)}}{g_{_{33}}(z)}g_{_{34}}(z)\Big).$\\\\

			\item [(vii)] Since $\mathtt{C}_{_{\theta_{1}}} \subseteq \mathtt{C}_{_{\theta_{3}}}, \mathtt{C}_{_{\theta_{1}}} \subseteq \mathtt{C}_{_{\theta_{4}}}$ for $\theta \in \{0,1,2\nu,3+2\nu\}$ and $\mathtt{C}_{_{\theta_{2}}} \subseteq \mathtt{C}_{_{\theta_{4}}}$ for $\theta \in \{0,3+2\nu\},$ it follows that 
			$g_{_{33}}(z)|g_{_{11}}(z),
			g_{_{44}}(z)|g_{_{11}}(z) ~~ \text{for } \theta \in \{0,1,2\nu,3+2\nu\}$ and
			$g_{_{44}}(z)|g_{_{22}}(z) ~~ \text{for } \theta \in \{0,3+2\nu\}.$\\
			
			Also, $k_{_{\theta}}\Big(2g_{_{22}}(z)+k_{_{\theta}} g_{_{23}}(z)+2k_{_{\theta}}g_{_{24}}(z)\Big)$ belongs to $\mathtt{C_{_{\theta}}},$ it follows that $2k_{_{\theta}} \big(g_{_{22}}(z)+g_{_{23}}(z)\big)$ belongs to $\mathtt{C_{_{\theta}}}$ for $\theta \in \{1,2\nu\},$ which implies that $\big(g_{_{22}}(z)+g_{_{23}}(z)\big)$ belongs to $\mathtt{C}_{_{\theta_{4}}}.$ Therefore, $g_{_{44}}(z)|g_{_{22}}(z)+g_{_{23}}(z)$ for $\theta \in \{1,2\nu\}.$\\
			
			\item [(viii)] Since $k_{_{\theta}}\Big(g_{_{11}}(z)+2g_{_{12}}(z)+k_{_{\theta}}g_{_{13}}(z)+2k_{_{\theta}}g_{_{14}}(z)\Big) -\dfrac{g_{_{11}}(z)}{g_{_{33}}(z)}\Big(k_{_{\theta}}g_{_{33}}(z)+2k_{_{\theta}}g_{_{34}}(z)\Big)$ belongs to $\mathtt{C}_{_{\theta}},$ it follows that
			$2k_{_{\theta}}\Big(g_{_{12}}(z)-\dfrac{g_{_{11}}(z)}{g_{_{33}}(z)}g_{_{34}}(z)\Big)+k_{_{\theta}}^{2}g_{_{13}}(z)+2k_{_{\theta}}^{2}g_{_{14}}(z)$ belongs to  $\mathtt{C}_{_{\theta}}.$ Therefore,\\
			$2k_{_{\theta}}\Big(g_{_{12}}(z)+g_{_{13}}(z)-\dfrac{g_{_{11}}(z)}{g_{_{33}}(z)}g_{_{34}}(z)\Big) \in \mathtt{C}_{_{\theta}}~ \text{for } \theta \in\{1,2\nu\}$ and\\
			$2k_{_{\theta}}\Big(g_{_{12}}(z)-\dfrac{g_{_{11}}(z)}{g_{_{33}}(z)}g_{_{34}}(z)\Big) \in \mathtt{C}_{_{\theta}}~ \text{for } \theta \in\{0,3+2\nu\}$\\
			which implies that\\
			$g_{_{12}}(z)+g_{_{13}}(z)-\dfrac{g_{_{11}}(z)}{g_{_{33}}(z)}g_{_{34}}(z) \in \mathtt{C}_{_{\theta_{4}}}~ \text{for } \theta \in\{ 1,2\nu\}$ and\\
			$g_{_{12}}(z)-\dfrac{g_{_{11}}(z)}{g_{_{33}}(z)}g_{_{34}}(z) \in \mathtt{C}_{_{\theta_{4}}}~ \text{for } \theta \in\{0,3+2\nu\}.$\\\\
			Hence,\\
			$g_{_{44}}(z)|g_{_{12}}(z)+g_{_{13}}(z)-\dfrac{g_{_{11}}(z)}{g_{_{33}}(z)}g_{_{34}}(z)~\text{for } \theta \in\{1,2\nu\}$ and\\
			$g_{_{44}}(z)|g_{_{12}}(z)-\dfrac{g_{_{11}}(z)}{g_{_{33}}(z)}g_{_{34}}(z) ~\text{for } \theta \in\{0,3+2\nu\}.$\\
			
			Also, $2k_{_{\theta}}\Big(g_{_{11}}(z)+2g_{_{12}}(z)+k_{_{\theta}}g_{_{13}}(z)+2k_{_{\theta}}g_{_{14}}(z)\Big) -2\dfrac{g_{_{11}}(z)}{g_{_{33}}(z)}\Big(k_{_{\theta}}g_{_{33}}(z)+2k_{_{\theta}}g_{_{34}}(z))\Big) \in \mathtt{C}_{_{\theta}}$ implies that $2k_{_{\theta}}^{2}g_{_{13}}(z)\in \mathtt{C}_{_{\theta}}.$ It follows that $2k_{_{\theta}}g_{_{13}}(z)\in \mathtt{C}_{_{\theta}}~ \text{for } \theta \in \{\nu,3\nu,2+\nu,2+3\nu\},$ and hence $g_{_{13}}(z)\in \mathtt{C}_{_{\theta_{4}}}$ for $\theta \in \{\nu,3\nu,2+\nu,2+3\nu\}.$ Thus 	$g_{_{44}}(z)|g_{_{13}}(z) ~ \text{for } \theta \in \{\nu,3\nu,2+\nu,2+3\nu\}.$ 
		\end{itemize} 
	\end{proof} 
	
	\section{Rank and Cardinality of cyclic codes of arbitrary length over $\mathtt{R}_{_{\theta}}, \theta \in \mathtt{S}$}
	In this section, the rank and cardinality of cyclic codes of arbitrary length over $\mathtt{R}_{_{\theta}}, \theta \in \mathtt{S},$ have been obtained by determining a minimal spanning set of a cyclic code over $\mathtt{R}_{_{\theta}}.$  \\
	
	\begin{theorem} Let $\mathtt{C}_{_{\theta}}=\langle g_{_{\theta_{1}}}(z),g_{_{\theta_{2}}}(z),g_{_{\theta_{3}}}(z),g_{_{\theta_{4}}}(z)\rangle$ be a cyclic code of arbitrary length $n$ over the ring $\mathtt{R}_{_{\theta}},\theta \in \mathtt{S}$, where the generators $g_{_{\theta_{1}}}(z)= g_{_{11}}(z)+2g_{_{12}}(z)+k_{_{\theta}}g_{_{13}}(z)+2k_{_{\theta}}g_{_{14}}(z)$, $g_{_{\theta_{2}}}(z)=2g_{_{22}}(z)+k_{_{\theta}}g_{_{23}}(z)+2k_{_{\theta}}g_{_{24}}(z)$, $g_{_{\theta_{3}}}(z)=k_{_{\theta}}g_{_{33}}(z)+2k_{_{\theta}}g_{_{34}}(z)$, $g_{_{\theta_{4}}}(z)=2k_{_{\theta}}g_{_{44}}(z)$ are in the unique form as given in Theorem 3.3. Then $rank(\mathtt{C}_{_{\theta}})$ is $n+s_{_{1}}+\tilde{s}-s_{_{2}}-s_{_{3}}-s_{_{4}},$
		where $s_{_{i}}=$ deg $g_{_{ii}}(z)$ for $1 \leq i \leq 4$ and $\tilde{s}=	min\{s_{_{2}},s_{_{3}}\}.$\\
		
	\end{theorem}
	\begin{proof}
		It can be easily seen that the set $\mathtt{A}_{_{\theta}}=\{g_{_{\theta_{1}}}(z),zg_{_{\theta_{1}}}(z),\cdots,z^{n-s_{_{1}}-1}g_{_{\theta_{1}}}(z),\\g_{_{\theta_{2}}}(z),zg_{_{\theta_{2}}}(z),\cdots,z^{n-s_{_{2}}-1}g_{_{\theta_{2}}}(z),g_{_{\theta_{3}}}(z),zg_{_{\theta_{3}}}(z), \cdots,z^{n-s_{_{3}}-1}g_{_{\theta_{3}}}(z),g_{_{\theta_{4}}}(z),zg_{_{\theta_{4}}}(z),\\\cdots,z^{n-s_{_{4}}-1}g_{_{\theta_{4}}}(z)\}$ is a spanning set of $\mathtt{C}_{_{\theta}}.$\\
		
		To prove that $rank~(\mathtt{C}_{_{\theta}})$ is $n+s_{_{1}}+\tilde{s}-s_{_{2}}-s_{_{3}}-s_{_{4}},$ it is sufficient to show that the set $\mathtt{B}_{_{\theta}}=\{ g_{_{\theta_{1}}}(z),zg_{_{\theta_{1}}}(z),\cdots,z^{n-s_{_{1}}-1}g_{_{{\theta}_{1}}}(z),g_{_{\theta_{2}}}(z),zg_{_{\theta_{2}}}(z),\cdots,z^{s_{_{1}}-s_{_{2}}-1}g_{_{\theta_{2}}}(z),g_{_{\theta_{3}}}(z),\\zg_{_{\theta_{3}}}(z), \cdots,z^{s_{_{1}}-s_{_{3}}-1}g_{_{\theta_{3}}}(z),g_{_{\theta_{4}}}(z),zg_{_{\theta_{4}}}(z),\cdots,z^{\tilde{s}-s_{_{4}}-1}g_{_{\theta_{4}}}(z)\}$ is a minimal spanning set of $\mathtt{C}_{_{\theta}},$ where $\tilde{s}=	min\{s_{_{2}},s_{_{3}}\}.$\\
		
		In order to prove that the set $\mathtt{B}_{_{\theta}}$ spans $\mathtt{C}_{_{\theta}},$ it is enough to show that $z^{\tilde{s}-s_{_{4}}}g_{_{\theta_{4}}}(z),\\ z^{s_{_1}-s_{_{3}}}g_{_{\theta_{3}}}(z), z^{s_{_1}-s_{_{2}}}g_{_{\theta_{2}}}(z) \in span({\mathtt{B}_{_{\theta}}})$. First, 
		let us suppose that $\tilde{s}=s_{_{3}}.$ As $g_{_{44}}(z)|g_{_{33}}(z)$ in $Z_{2}[z]/\left\langle z^{n}-1 \right\rangle,$ there exists some $m(z) \in Z_{2}[z]$ with deg $m(z) =s_{_{3}}-s_{_{4}}$ such that $g_{_{33}}(z)=g_{_{44}}(z)m(z)$ =$g_{_{44}}(z)\big(m_{_{0}}+zm_{_{1}}+\cdots+z^{s_{_{3}}-s_{_{4}}-1}m_{{s_{_{3}}}-s_{_{4}}-1}+z^{s_{_{3}}-s_{_{4}}}\big), m_{_{i}} \in Z_{2}.$
		Multiplying both sides by $2k_{_{\theta}}$, we get
		$$2g_{_{\theta_{3}}}(z)=\big(m_{_{0}}+zm_{_{1}}+\cdots+z^{s_{_{3}}-s_{_{4}}-1}m_{{s_{_{3}}}-s_{_{4}}-1}\big)g_{_{\theta_{4}}}(z)+z^{s_{_{3}}-s_{_{4}}}g_{_{\theta_{4}}}(z)$$ which implies that $z^{s_{_{3}}-s_{_{4}}}g_{_{\theta_{4}}}(z) \in span(\mathtt{B}_{_{\theta}}).$
		Next, suppose that $\tilde{s}=s_{_{2}}.$
		Using the divisibilties $g_{_{44}}(z)|g_{_{22}}(z)$ for $\theta \in \{0,3+2\nu\}, g_{_{44}}(z)|g_{_{22}}(z)+g_{_{23}}(z)$ for  $\theta \in \{1,2\nu\}$ and $g_{_{44}}(z)|g_{_{23}}(z)$ for $\theta \in \{\nu,3\nu,2+\nu,2+3\nu\}$, it can be proved that $z^{s_{_{2}}-s_{_{4}}}g_{_{\theta_{4}}}(z) \in span(\mathtt{B}_{_{\theta}})$ by working on the same lines as above. Thus, we have $z^{\tilde{s}-s_{_{4}}}g_{_{\theta_{4}}}(z) \in span(\mathtt{B}_{_{\theta}}),$ where $\tilde{s}=min\{s_{_{2}},s_{_{3}}\}$.\\
		
		Now, we proceed to prove that $z^{s_{_{1}}-s_{_{3}}}g_{_{\theta_{3}}}(z) \in span(\mathtt{B}_{_{\theta}}).$ Since deg $z^{s_{_{1}}-s_{_{3}}} g_{_{\theta_{3}}}(z) =$ deg $g_{_{\theta_{1}}}(z)=s_{_{1}},$ there exist a polynomial $r_{_{1}}(z)$ such that 
		\begin{equation}
			r_{_{1}}(z)=z^{s_{_{1}}-s_{_{3}}}g_{_{\theta_{3}}}(z)-k_{_{\theta}}g_{_{\theta_{1}}}(z).
		\end{equation}
		Clearly, $r_{_{1}}(z) \in \mathtt{C_{_{\theta}}}.$ Moreover, either $r_{_{1}}(z)=0$ or  deg $r_{_{1}}(z)< s_{_{1}}.$ If $r_{_{1}}(z)=0,$ then  $z^{s_{_{1}}-s_{_{3}}}g_{_{\theta_{3}}}(z) \in span(\mathtt{B_{_{\theta}}}).$ If deg $
		r_{_{1}}(z) < s_{_{1}},$ then it is easy to see that $r_{_{1}}(z)$ is of the type $g_{_{\theta_{3}}}(z)$ or $g_{_{\theta_{4}}}(z).$\\
		
		If $r_{_{1}}(z)$ is of the type $g_{_{\theta_{4}}}(z),$ then due to the minimality of degree of $g_{_{\theta_{4}}}(z),$ we have deg $r_{_{1}}(z) \geq s_{_{4}}.$ Therefore, there exist a polynomial $r_{_{2}}(z)$ such that 
		\begin{equation*}
			r_{_{2}}(z)=r_{_{1}}(z)-z^{{\text{ deg }r_{_{1}}(z)-s_{_{4}}}}g_{_{\theta_{4}}}(z).
		\end{equation*}
		It is easy to see that $r_{_{2}}(z) \in \mathtt{C_{_{\theta}}}$ and it is of the type $g_{_{\theta_{_4}}}(z).$ Also, either $r_{_{2}}(z)=0$ or deg $ r_{_{2}}(z) <$ deg $r_{_{1}}(z).$ If $r_{_{2}}(z)=0,$ then $r_{_{1}}(z)=z^{{\text{ deg }r_{_{1}}(z)-s_{_{4}}}}g_{_{\theta_{4}}}(z).$ Subsituting the value of $r_{_{1}}(z)$ in (4.1), we see that $z^{s_{_{1}}-s_{_{3}}} g_{_{\theta_{3}}}(z) \in span(\mathtt{B_{_{\theta}}}).$ If deg $ r_{_{2}}(z) <$ deg $r_{_{1}}(z),$ then after repeating the argument a finite number of times we obtain a polynomial $r_{_{l}}(z) = r_{_{l-1}}(z)- z^{{\text{ deg }r_{_{l-1}}(z)-s_{_{4}}}}g_{_{\theta_{4}}}(z)$ such that $r_{_{l}}(z) \in \mathtt{C_{_{\theta}}}$ and it is of the type $g_{_{\theta_{_4}}}(z).$ Moreover, $r_{_{l}}(z)=0$ or deg $r_{_{l}}(z) < s_{_{4}}.$ Since $r_{_{l}}(z)$ is of the type $g_{_{\theta_{_4}}}(z),$ deg $r_{_{l}}(z)$ cannot be less than $s_{_{4}}.$ Therefore, $r_{_{l}}(z)=0.$ Hence, from equation (4.1), we have,\\
		$z^{s_{_{1}}-s_{_{3}}} g_{_{\theta_{3}}}(z)=k_{_{\theta}}g_{_{\theta_{1}}}(z)+r_{_{1}}(z)=k_{_{\theta}}g_{_{\theta_{1}}}(z)+z^{{\text{ deg }r_{_{1}}(z)-s_{_{4}}}}g_{_{\theta_{4}}}(z)+r_{_{2}}(z)\\=k_{_{\theta}}g_{_{\theta_{1}}}(z)+z^{{\text{ deg }r_{_{1}}(z)-s_{_{4}}}}g_{_{\theta_{4}}}(z)+z^{{\text{ deg }r_{_{2}}(z)-s_{_{4}}}}g_{_{\theta_{4}}}(z)+\cdots+z^{{\text{ deg }r_{_{l-1}}(z)-s_{_{4}}}}g_{_{\theta_{4}}}(z).$\\
		
		It follows that $z^{s_{_{1}}-s_{_{3}}} g_{_{\theta_{3}}}(z) \in span(\mathtt{B_{_{\theta}}}),$ in case $r_{_{1}}(z)$ is of the type $g_{_{\theta_{4}}}(z).$ A simiar arguments can be used to prove that $z^{s_{_{1}}-s_{_{3}}} g_{_{\theta_{3}}}(z) \in span(\mathtt{B_{_{\theta}}})$ in case $r_{_{1}}(z)$ is of the type $g_{_{\theta_{_3}}}(z).$\\
		By using a similar argument as above, it can be proved that $z^{s_{_{1}}-s_{_{2}}}g_{_{\theta_{2}}}(z) \in span(\mathtt{B_{_{\theta}}}).$ Thus, $\mathtt{B_{_{\theta}}}$ is a spanning set of $\mathtt{C_{_{\theta}}}.$\\ 
		
		To prove that the set $\mathtt{B_{_{\theta}}}$ is a minimal spanning set, it is enough to show that none of $z^{n-s_{_{1}}-1}g_{_{\theta_{1}}}(z), z^{s_{_{1}}-s_{_{2}}-1}g_{_{\theta_{2}}}(z),
		z^{s_{_{1}}-s_{_{3}}-1}g_{_{\theta_{3}}}(z)$ and $z^{\tilde{s}-s_{_{4}}-1}g_{_{\theta_{4}}}(z)$ can be written as a linear combination of other elements of $\mathtt{B}_{_{\theta}}.$ Suppose, if possible, that $z^{n-s_{_{1}}-1}g_{_{\theta_{1}}}(z)$ can be written as a linear combinations of other elements of $\mathtt{B}_{_{\theta}},$ i.e, 
		\begin{equation}
			z^{n-s_{_{1}}-1}g_{_{\theta_{1}}}(z)= a(z)g_{_{\theta_{1}}}(z)
			+b(z)g_{_{\theta_{2}}}(z)
			+c(z)g_{_{\theta_{3}}}(z)+d(z)g_{_{\theta_{4}}}(z),
		\end{equation}
		where deg $a(z) < n-s_{_{1}}-1$, deg $b(z) < s_{_{1}}-s_{_{2}},$ deg $c(z) < s_{_{1}}-s_{_{3}}$ and deg $d(z) < \tilde{s}-s_{_{4}}.$ On multiplying equation (4.2) on both sides by $2k_{_{\theta}}$ for $\theta \in \{0,1,2\nu,3+2\nu\},$ we get
		\begin{equation}
			2k_{_{\theta}}z^{n-s_{_{1}}-1} g_{_{11}}(z)= 2k_{_{\theta}}a(z)g_{_{11}}(z), ~\theta \in \{0,1,2\nu,3+2\nu\}.
		\end{equation}
		On multiplying equation (4.2) on both sides by $2(k_{_{\theta}}-1)$ for $\theta \in \{\nu,3\nu,2+\nu,2+3\nu\},$ we get
		\begin{equation}
			2(k_{_{\theta}}-1)z^{n-s_{_{1}}-1} g_{_{11}}(z)= 2(k_{_{\theta}}-1)a(z)g_{_{11}}(z), ~\theta \in \{\nu,3\nu,2+\nu,2+3\nu\}.
		\end{equation}
		The equations (4.3) and (4.4) are not possible as degrees of left hand side and right hand side in each of these equations do not match. Thus, $ 	z^{n-s_{_{1}}-1}g_{_{\theta_{1}}}(z)$ can not be written as a linear combination of other elements of $\mathtt{B_{_{\theta}}}.$ Using a similar argument, it can be shown that none of $z^{s_{_{1}}-s_{_{2}}-1}g_{_{\theta_{2}}}(z), z^{s_{_{1}}-s_{_{3}}-1}g_{_{\theta_{3}}}(z)$ and $z^{\tilde{s}-s_{_{4}}-1}g_{_{\theta_{4}}}(z)$ can be written as a linear combination of other elements of $\mathtt{B}_{_{\theta}}.$ Hence, $\mathtt{B}_{_{\theta}}$ is a minimal spanning set of $\mathtt{C}_{_{\theta}}.$ \\
		Further,  $rank(\mathtt{C}_{_{\theta}})=$ Number of elements in $\mathtt{B}_{_{\theta}}=(n-s_{_1})+(s_{_{1}}-s_{_{2}})+(s_{_{1}}-s_{_{3}})+(\tilde{s}-s_{_{4}})=n+s_{_{1}}+\tilde{s}-s_{_{2}}-s_{_{3}}-s_{_{4}},$ where $\tilde{s}= min\{s_{_{2}},s_{_{3}}\}.$
	\end{proof}
	Corollary 1 below follows immediately from the above theorem.\\
	
	\textbf{Corollary 1} Let $\mathtt{C}_{_{\theta}}=\langle g_{_{\theta_{1}}}(z),g_{_{\theta_{2}}}(z),g_{_{\theta_{3}}}(z),g_{_{\theta_{4}}}(z)\rangle$ be a cyclic code of arbitrary length $n$ over the ring $\mathtt{R}_{_{\theta}},\theta \in \mathtt{S}$, where the generators $g_{_{\theta_{1}}}(z)= g_{_{11}}(z)+2g_{_{12}}(z)+k_{_{\theta}}g_{_{13}}(z)+2k_{_{\theta}}g_{_{14}}(z)$, $g_{_{\theta_{2}}}(z)=2g_{_{22}}(z)+k_{_{\theta}}g_{_{23}}(z)+2k_{_{\theta}}g_{_{24}}(z)$, $g_{_{\theta_{3}}}(z)=k_{_{\theta}}g_{_{33}}(z)+2k_{_{\theta}}g_{_{34}}(z)$, $g_{_{\theta_{4}}}(z)=2k_{_{\theta}}g_{_{44}}(z).$ Then Cardinality of $\mathtt{C}_{_{\theta}}$ is \[ |\mathtt{C_{_{\theta}}}|= \begin{cases}
		2^{4n+s_{_{1}}+\tilde{s}-3s_{_{2}}-2s_{_{3}}-s_{_{4}}} & ;g_{_{23}}(z) \ne 0 \\
		2^{4n+\tilde{s}-2s_{_{2}}-2s_{_{3}}-s_{_{4}}} & ;g_{_{23}}(z) =0 \\
	\end{cases}, \] where $s_{_{i}}=$ deg $g_{_{ii}}(z)$ for $1 \leq i \leq 4$ and $\tilde{s} =min \{s_{_{2}},s_{_{3}}\}$.\\

	The following examples iilustrate some of our results.\\
	
	\begin{example} Let $\mathtt{C}_{_{\theta}} =\langle z^{3}+z^{2}+z+1+\nu(z+3), 2(z^{2}+1)+2\nu, \nu(z^{2}+1), 2\nu(z+1)\rangle$ be a cyclic code of length $4$ over the ring $\mathtt{R}_{_{\theta}}$ for $\theta =2\nu.$ Here $s_{_{1}}=3, s_{_{2}}=2, s_{_{3}}=2, s_{_{4}}=1.$ Using Theorem 4.1, minimal spanning set of $\mathtt{C}_{_{\theta}}$ is $\{z^{3}+z^{2}+z+1+\nu(z+3), 2(z^{2}+1)+2\nu, \nu(z^{2}+1), 2\nu(z+1)\}.$ Hence rank($\mathtt{C}_{_{\theta}})=4$ and $|\mathtt{C_{_{\theta}}}|=2^{9}.$
	\end{example}
	
	\begin{example}
		Let $\mathtt{C}_{_{\theta}} = \langle z^{3}+z^{2}+z+1 +(1+\nu), 2(z^{2}+1), (1+\nu)(z+1), 2(1+\nu)\rangle$ be a cyclic code of length $4$ over the ring $\mathtt{R}_{_{\theta}}$ for $\theta=3+2\nu.$ Here $s_{_{1}}=3, s_{_{2}}=2, s_{_{3}}=1, s_{_{4}}=0.$ Using Theorem 4.1, we have minimal spanning set of $\mathtt{C}_{_{\theta}}$ is $\{z^{3}+z^{2}+z+1 +(1+\nu), 2(z^{2}+1), (1+\nu)(z+1),z(1+\nu)(z+1), 2(1+\nu) \}.$ Hence rank($\mathtt{C}_{_{\theta}})=5$ and $|\mathtt{C_{_{\theta}}}|=2^{11}.$ 
	\end{example}
	
	\begin{example} Let $\mathtt{C}_{_{\theta}} =\langle z^{5}+z^{4}+z^{3}+z^{2}+z+1 +\nu(z^{4}+z^{2}+1),2(z+1)+\nu(z+1),\nu(z^{5}+z^{4}+z^{3}+z^{2}+z+1), 2\nu \rangle$ be a cyclic code of length $6$ over the ring $\mathtt{R}_{_{\theta}}$ for $\theta=\nu.$ Here $s_{_{1}}=5, s_{_{2}}=1, s_{_{3}}=5, s_{_{4}}=0.$ Using Theorem 4.1, minimal spanning set of $\mathtt{C}_{_{\theta}}$ is $\{z^{5}+z^{4}+z^{3}+z^{2}+z+1 +\nu(z^{4}+z^{2}+1),2(z+1)+\nu(z+1),2z(z+1)+\nu z(z+1),2z^{2}(z+1)+\nu z^{2}(z+1),2z^{3}(z+1)+\nu z^{3}(z+1),2\nu\}.$ Hence rank($\mathtt{C}_{_{\theta}})=6$ and $|\mathtt{C_{_{\theta}}}|=2^{17}.$
	\end{example}
	
	\begin{example} Let $\mathtt{C}_{_{\theta}} =\langle z^{5}+z^{4}+z^{3}+z^{2}+z+1 +\nu(z^{2}+z+1)+2\nu z, 2(z^{4}+z^{2}+1), \nu(z^{3}+3),2\nu(z^{2}+z+1)\rangle$ be a cyclic code of length $6$ over the ring $\mathtt{R}_{_{\theta}}$ for $\theta=0.$ Here $s_{_{1}}=5, s_{_{2}}=4, s_{_{3}}=3, s_{_{4}}=2.$ Using Theorem 4.1, minimal spanning set of $\mathtt{C}_{_{\theta}}$ is $\{z^{5}+z^{4}+z^{3}+z^{2}+z+1 +\nu(z^{2}+z+1)+2\nu z, 2(z^{4}+z^{2}+1), \nu(z^{3}+3),z\nu(z^{3}+3), 2\nu(z^{2}+z+1)\}.$ Hence rank($\mathtt{C}_{_{\theta}})=5$ and $|\mathtt{C_{_{\theta}}}|=2^{11}.$
	\end{example}
	
	\section{Conclusion}
	In this paper, the structure of cyclic codes of arbitrary length over the rings $Z_{4}+\nu Z_{4}$ for those values of $\nu^{2}$ for which these are non-chain rings has been established. A unique form of the generators of these codes has been obtained. Further, formulae for  rank and cardinality of these codes have been established by finding minimal spanning sets for these codes.

\end{document}